# Electronic Structures and Their Landau Quantizations in Twisted Graphene Bilayer and Trilayer

Long-Jing Yin[§], Jia-Bin Qiao[§], Wei Yan, Rui Xu, Rui-Fen Dou, Jia-Cai Nie, and Lin He*

**Electronic structures and their Landau quantizations in twisted graphene bilayer and trilayer are investigated using scanning tunnelling microscopy and spectroscopy. In the twisted trilayer, the top graphene layer and second layer are *AB* (Bernal) stacking and there is a stacking misorientation between the second layer and third layer. Both the twisted bilayer and trilayer exhibit two pronounced low-energy van Hove singularities (VHSs) in their spectra. Below the VHSs, the observed Landau level quantization in the twisted bilayer is identical to that of massless Dirac fermion in graphene monolayer. Our result demonstrates that both the VHSs and Fermi velocity of the twisted bilayer depends remarkably on the twist angle and the interlayer coupling strength. In the twisted trilayer, we directly observe Landau quantization of massive Dirac fermion with a sizable band gap 105 ± 5 meV, which results in valley (layer) polarization of the lowest Landau levels. Such a result is similar to the expected Landau quantization in Bernal graphene bilayer with a moderate electric field.**

Department of Physics, Beijing Normal University, Beijing, 100875, People's Republic of China
[§]These authors contributed equally to this work.
* Email: helin@bnu.edu.cn

The electronic properties of graphene bilayer and trilayer depend sensitively on their stacking order[1-18]. The two common stacking configurations of bilayer, *i.e.*, the AB (Bernal) and AA-stacked bilayer[1], and the two natural stable trilayer, *i.e.*, the ABA and ABC-stacked trilayer[3], exhibit very different electronic properties even though there is only subtle distinction between their stacking orders. As an example, a shift of one graphene layer by the carbon-carbon bond distance could lead to a transition between the AB bilayer (ABC trilayer) and the AA bilayer (ABA trilayer)[5]. Very recently, the introduction of a stacking misorientation has further expanded the allotropes of the graphene bilayer and trilayer dramatically and, more importantly, these resultants have been demonstrated to show strong twist-dependent electronic spectra and properties[2,15-17,19-29]. Therefore, the twisted angle is treated as an unique degree of freedom to tune the electronic properties of graphene multilayer.

In this paper, we perform scanning tunnelling microscopy (STM) and spectroscopy (STS) measurements of twisted graphene bilayer and trilayer on highly oriented pyrolytic graphite (HOPG) surface. The twisted bilayer and trilayer regions are separated by a tilt grain boundary and both of them have the same twisted angle. The spatial evolution of low-energy van Hove singularities (VHSs), which originate from the two saddle points in the band structures of the twisted graphene bilayer and trilayer[2,15-17], along a trajectory traversing the boundary is recorded. More importantly, we measure the Landau quantizations in both the twisted bilayer and trilayer, which are rarely explored in experiment so far.

The STM system was an ultrahigh vacuum scanning probe microscope (USM-1500) from UNISOKU (see Supplementary Information for experimental method). The surface few-layer of HOPG usually decouples from the bulk[2,29-32], therefore, the surface of HOPG provides a natural ideal platform to probe the electronic spectra of graphene layers. Figure 1a-c shows typical STM images of a twisted graphene bilayer and trilayer on a HOPG surface. The twisted bilayer and trilayer regions are separated by a tilt grain boundary of the second graphene layer, as schematically shown in Fig. 1d. The results of Fig. 1e and Fig. 1f indicate that the topmost graphene layer is continuous. The twisted angle $\theta$ between neighboring graphene layers can be obtained from the period of the Moiré patterns $D$, using $D = a/[2 \sin(\theta/2)]$ with the graphene lattice constant a ~ 0.246 nm[2,15-17,21-23,27-29]. Here, we obtain $\theta$ ~ 2.8° in both the twisted bilayer (region I, denoted as TB) and the twisted trilayer (region II, denoted as ABT), which consists well with the result derived from the Fourier transforms of the STM image (see Fig. S1 of Supplementary Information for more experimental result and discussion about the possible origin of the same twisted angle in the two regions). In the twisted trilayer, the top graphene layer and second layer are AB stacking, as demonstrated by the observed triangular lattice in Fig. 1e, and the Moiré pattern in region II arises from the stacking misorientation between the second layer and third layer, which results in a lower contrast of the Moiré pattern in the STM image. According to a previous paper[29], the emergence of such a unique system is usually accompanied by the appearance of a tilt grain boundary. This is further confirmed in this work (see Fig. S2 of Supplementary Information for more experimental data).

To investigate the influence of the stacking order and layer number on the electronic spectra, we measure the spatial evolution of tunnelling spectra along a line across the tilt grain boundary [Fig. 2a]. As shown in Fig. 2b, the spectroscopy shows an abrupt change across the boundary, but it smoothly evolves from TB to ABT over a spatial extent of about 2.4 nm because of the influence of the boundary. In both the TB and ABT regions, the spectra exhibit the two low-energy VHSs[2,15-17]. However, the energy difference of the VHSs, $\Delta E_{VHS}$, abruptly reduces from the TB (~ 0.54 eV) to the ABT (~ 0.44 eV), as shown in Fig. 2c. Such a reduction of the $\Delta E_{VHS}$ is a direct result of introducing a third layer on top of a twisted bilayer[29], as shown in Fig. 2d. With identical twisted angle and interlayer coupling, the energy difference of the saddle points in the electronic band structures of the ABT is much smaller than that of the TB.

Besides the reduction of the $\Delta E_{VHS}$, the spectra recorded in the ABT region at positions away from the boundary (the distance away from the boundary is larger than 2.4 nm) show another two peaks within the two VHSs, as shown in Fig. 2b (see Fig. S3 of Supplementary Information for more experimental data). In the ABT, the topmost two layers are AB-stacked. The substrate (the HOPG here) breaks the symmetry of the two layers. Such an effect can be treated as an effective external electric field applying on AB-stacked bilayer, which consequently generates a finite gap in the parabolic bands of the ABT (Fig. 2d)[1,4,6,33,34]. Two peaks in the spectrum are signatures of the density of states (DOS) peaks generated at the conduction band and valence band edges. Then, the energy gap in the parabolic bands of the ABT is estimated as 105 ± 5 meV (the effective electric field on the AB-stacked bilayer is estimated to be about 100 meV, see Fig. S4 of Supplementary Information). In the TB region, the effective electric field between the adjacent two layers leads to relative move of the two Dirac points in opposite directions in energy[2,15], as shown in Fig. 2d. This results in an obvious asymmetry of the two VHSs. Additionally, the energy states at the two saddle points have different weights in the two layers in such a case[2]. As a consequence, we observe a pronounced asymmetry between the intensity of the two VHSs since that the STS spectrum mainly reflects the local DOS of the surface layer at the position of the STM tip.

We further perform STS measurements for different magnetic fields of the two systems. Because of their different electronic spectra, it is expected to observe distinct Landau quantization in the TB and ABT. Figure 3 displays our experimental results and analysis of the TB. In the TB region, it is expected to generate two series of Landau level (LL) sequences, which are offset from each other in energy, below the VHSs in high magnetic fields. Each of the LL sequence depends on the square-root of both level index $n$ and magnetic field $B$[25,31,35]:

$$E_n = E_D + \text{sgn}(n)\sqrt{2e\hbar v_F^2 |n| B}, \qquad n = 0, \pm 1, \pm 2, \ldots \qquad (1)$$

Here $E_D$ is the energy of Dirac point, $e$ is the electron charge, and $v_F$ is the Fermi velocity. The sequence of LLs shown in Fig. 3a and 3b is unique to massless Dirac fermion and is expected to be observed in the TB[22]. In the experiment, we only observe one LLs sequence since that the quasiparticles of the topmost graphene layer are mainly coming from one of the two Dirac cones, as shown in the inset of Fig. 3b, and the STM probes predominantly the top layer.

In the literature, the theoretical calculations predicted pronounced Fermi velocity renormalization in the TB with small twisted angles and the Fermi velocity in the TB with $\theta \sim 2.8°$ was predicted to be reduced to about 65% of that of a graphene monolayer[15,22,36]. However, a reasonable linear fitting of the experimental data to Eq. (1), as shown in Fig. 3b, is found yielding the Fermi velocity $v_F = (1.03 \pm 0.02) \times 10^6$ m/s. There is an obvious deviation between our experimental data and the theoretical result. Currently there is a lively discussion concerning the Fermi velocity renormalization in the TB and several contradicted experimental results are reported[16,22,25,37]. The strongest experimental evidence for the Fermi velocity renormalization is the observed Fermi velocity $v_F = 0.87 \times 10^6$ m/s, basing on Landau-level spectroscopy, in the TB with $\theta \sim 3.5°$ on a graphite surface[22]. Here we should point out that the pronounced Fermi velocity renormalization in the TB was predicted basing on a strong interlayer coupling ($t_\theta \sim 110$ meV). Theoretically, when the interlayer hopping is relative weak, we still can obtain the main features of the low-energy band structure of the TB, *i.e.*, the two displaced Dirac cones and the two VHSs, but without the significant reduction of the Fermi velocity, as shown in Fig. 4a. Therefore, we attribute the deviation between our experimental result and that reported in Ref. 22 mainly to the different interlayer hopping of the samples. In the TB with $\theta \sim 2.8°$, the interlayer hopping $t_\theta$ is estimated to be about 50 meV by comparing the measured $\Delta E_{\text{VHS}}$ with the theoretical result, as shown in Fig. 4a (it also can be roughly estimated according to $\Delta E_{\text{VHS}} \sim \hbar v_F \Delta K - 2t_\theta$). In the TB with $\theta \sim 3.5°$, as reported in Ref. 22, $t_\theta$ is estimated to be 110 meV. Therefore, we observed a negligible renormalization of the Fermi velocity in the TB, whereas the authors of Ref. 22 reported a pronounced reduction of the Fermi velocity.

To further study the effect of the interlayer coupling on the Fermi velocity of TB, we plot the predicted angle dependence of the renormalization for different interlayer hopping in Fig. 4b. Obviously, when $t_\theta = 50$ meV, there is only a slight reduction of the Fermi velocity for the TB with $\theta > 2.5°$, which is consistent well with our experimental result. Figure 4b also shows the Fermi velocities of other TB with different twist angles (see Fig. S5-7 of Supplementary Information for more experimental data). The Fermi velocity does not decrease monotonously with decreasing the twist angle, indicating that it is not only determined by the twist angle. Carefully examining the interlayer hopping of these samples reveals that the coupling strength between the two layers plays an important role in affecting the Fermi velocity. The measured Fermi velocity agrees well with the calculated result by taking into account the interlayer hopping of the TB (which is estimated according to the measured $\Delta E_{\text{VHS}}$), as shown in Fig. 4b. Our result in Fig. 4b also indicates that there is a large range of values for the interlayer coupling in different TB. The interlayer hopping of adjacent bilayer depends, in the simplest approximation, exponentially on the interlayer distance. In the TB, the stacking fault, tilt grain boundary, and roughness of substrate may affect the interlayer distance and stabilize it at various equilibrium distances.

In the ABT region, the observed LL spectra, as shown in Fig. 5, distinct from that in the TB region and follow that of massive Dirac fermion[38-42]

$$E_n = \pm [\hbar\omega_c(n(n-1))^{1/2} + E_g/2], \qquad n = 0,1,2\ldots \qquad (2)$$

Here $\omega_c = eB/m^*$ is the cyclotron frequency, $m^*$ is the effective mass of quasiparticles, and $E_g$ is the band gap. Such a result is reasonable since that the massive Dirac fermions of the ABT are localized at the topmost AB-stacked bilayer, as shown in Fig. S8 (see Supplementary Information for details of calculation). In high magnetic fields, two valley-polarized (layer-polarized) fourfold degenerate LLs are expected to be generated at the conduction and valence band edges with their positions independent of magnetic field[39]. This is demonstrated explicitly in our experiment, as shown in Fig. 5a. From the slope of the fitting curves in Fig. 5b, the effective mass of the electrons and holes localized at the topmost AB-stacked bilayer of the ABT is estimated to be about $(0.021 \pm 0.002)\, m_e$ and $(0.020 \pm 0.001)\, m_e$, respectively.

In summary, we address the electronic spectra and their Landau quantizations in the TB and the ABT. Our results demonstrate that a small twist could make a graphene monolayer and bilayer decouple from the substrate (the graphite here) sufficiently for electronic energy below the VHSs. This finding may have broad implications in tuning electronic properties of other two-dimensional van der Waals structures.

**Acknowledgments**

We thank A. K. Geim and E. Y. Andrei for helpful discussions. This work was supported by the National Basic Research Program of China (Grants Nos. 2014CB920903, 2013CBA01603, 2013CB921701), the National Natural Science Foundation of China (Grant Nos. 11422430, 11374035, 11474022, 51172029, 91121012), the program for New Century Excellent Talents in University of the Ministry of Education of China (Grant No. NCET-13-0054), Beijing Higher Education Young Elite Teacher Project (Grant No. YETP0238).

**Author contributions**

L.J.Y. and Y.W. performed the STM experiments. J.B.Q. and L.J.Y. analyzed the data and performed the theoretical calculations. L.H. conceived and provided advice on the experiment, analysis, and theoretical calculation. L.H., L.J.Y. and J.B.Q. wrote the paper. All authors participated in the data discussion.

**Competing financial interests:** The authors declare no competing financial interests.

**Figure Legends**

**Figure 1 | STM images of twisted bilayer and trilayer separated by a tilt grain boundary. a,** STM image of twisted graphene layers on graphite surface with two sets of Moiré superlattices showing a region of higher contrast (region I) and a region of lower contrast (region II). The two regions are separated by a tilt grain boundary, as indicated by the black dashed line. The periods of both the moiré patterns are almost the same, ~ 5.0 nm. Inset (bottom): a profile line across the boundary. Inset (top): Fourier transforms of the STM image with the six bright spots generated by the moiré superlattices around the boundary. **b,** STM image in region I of panel **a** showing moiré superlattices of twisted bilayer. **c,** Moiré superlattices of AB-twisted trilayer, as shown in region II of panel **a**. Insets of **b** and **c** are height profiles along the white dash lines. **d,** Schematic picture of the structure in panel **a**. **e,** High resolution current image of the white frame in panel **a**. The topmost graphene layer is continuous. We observe clear hexagonal lattice in region I and triangular lattice in region II. **f,** Fourier transforms of panel **a**. The outer six bright spots represent the reciprocal lattice of the topmost graphene. The middle six spots are related to the intervalley scattering generated by the boundary.

**Figure 2 | STS of the TB and ABT. a,** A STM image obtained around the boundary of the TB and ABT regions. **b,** Tunnelling spectroscopy as a function of tip position measured along the black arrow in **a**. The black dotted lines indicate the positions of VHSs in the TB and ABT regions. The three spectra overlaid onto the image, from the top to the bottom, are typical data obtained in the TB, near the boundary, and in the

ABT, respectively. Two peaks (labeled by black and red arrows) in the spectra of the ABT region are attributed to DOS peaks generated by the conduction band (black) and valence band (red) edges of Bernal graphene bilayer in an electric field. **c,** $\Delta E_{VHS}$ as a function of position around the boundary. The black and red dashed lines are the average values in the TB region (~ 0.54 eV) in the ABT region (~ 0.44 eV), respectively. **d,** Low-energy band structures for TB and ABT with no electric field and with a moderately sized electric field. In both cases, the energy difference of the VHSs in the ABT is much reduced with respect to that in the TB. An electric field (U ≠ 0) opens a gap $E_g$ in the parabolic bands of the ABT and leads to relative move of the two Dirac points of the TB in opposite directions in energy.

**Figure 3 | High magnetic field STS in the TB. a,** $dI/dV$-$V$ spectra taken in the TB region for different magnetic fields. The peaks are labeled with sequential LL indices of massless Dirac fermion. **b,** LL peak energies for applied fields of 6 T and 7 T obtained in panel **a** showing the square-root dependence on level index and magnetic field. The solid line is a linear fit of the data with the slope yielding a Fermi velocity of $v_F = (1.03 \pm 0.02) \times 10^6$ m/s. The inset shows schematic of low-energy dispersion with quantized LLs in TB. Only the LLs plotted in solid curves can be observed in the STS spectra.

**Figure 4 | Fermi velocity in the TB with various twist angles. a,** DOS of TB with interlayer hopping $t_\theta$ = 50 meV (red curve) and 110 meV (black curve). The inset

shows the corresponding low-energy band structure. There is a pronounced asymmetry between the two VHSs. From our calculation, both the $\Delta E_{VHS}$ and the Fermi velocity in the TB depend on the interlayer coupling strength. **b,** The solid squares are our experimental data, the open squares are the experimental data reported in Ref. 22. The curves are the predicted Fermi velocity renormalization with different interlayer hopping, which is estimated according to the measured $\Delta E_{VHS}$. Here the Fermi velocity is normalized with respect to $V_F = 1.10 \times 10^6$ m/s.

**Figure 5 | High magnetic field STS in the ABT. a,** $dI/dV$-$V$ spectra taken in the ABT region for different magnetic fields. The peaks marked by the index $n$ corresponding to the LLs of massive Dirac fermion. The eightfold degenerate zero-energy LL (4 for electrons and 4 for holes) was split into two valley-polarized quartets, as indicated by the dashed lines. The $LL_{(0,1,+)}$ and $LL_{(0,1,-)}$ projected on the top layer and the underlayer graphene, respectively (here +/- are valley index). The band gap is estimated to be $(105 \pm 5)$ meV. **b,** LL peak energies obtained in panel **a** plotted against $+(n(n-1))^{1/2}B$ for conduction band and $-(n(n-1))^{1/2}B$ for valence band. The red lines are the linear fit with the data in the electrons region and the holes region. The inset shows schematic low-energy dispersion of ABT with quantized LLs. Only the LLs plotted in solid curves can be observed in the STS spectra.

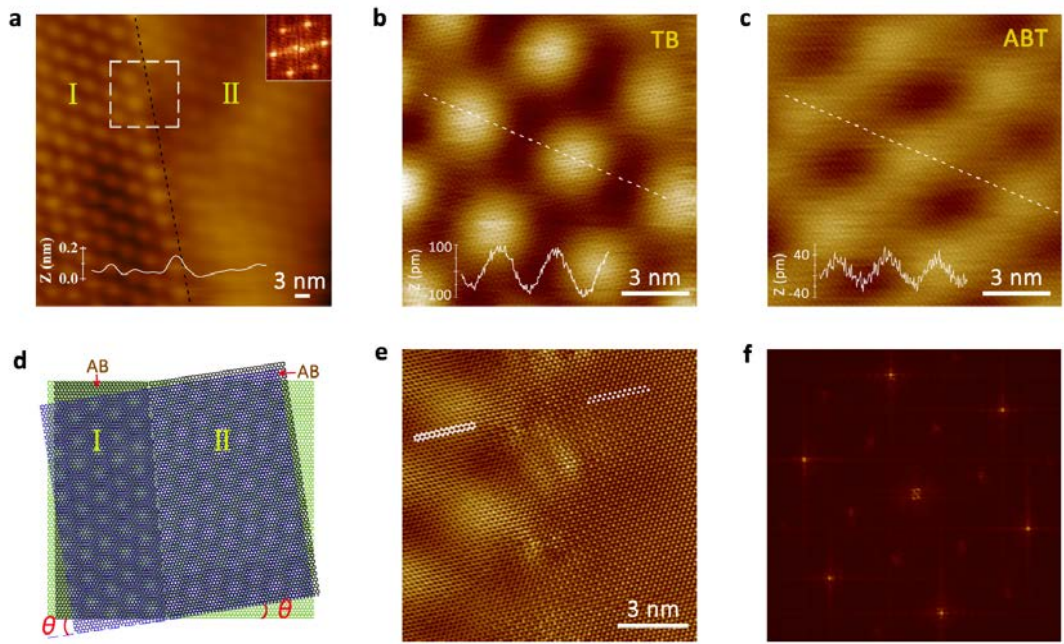

Figure 1

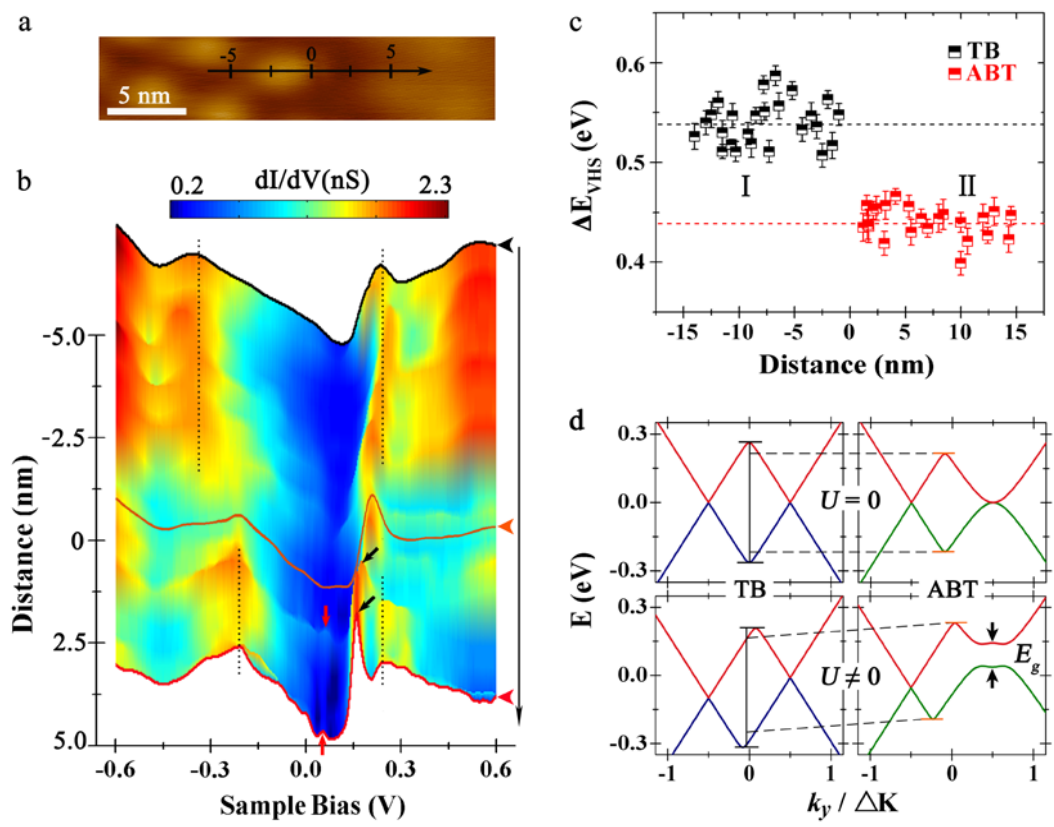

Figure 2

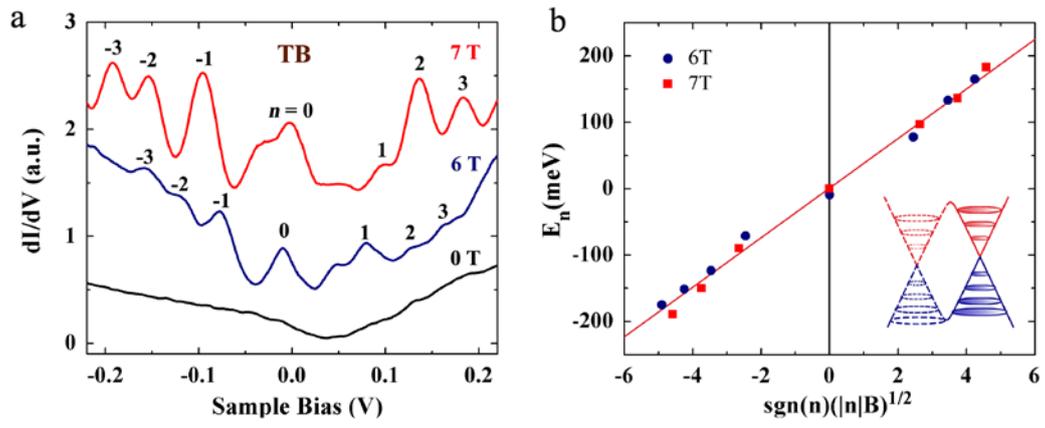

Figure 3

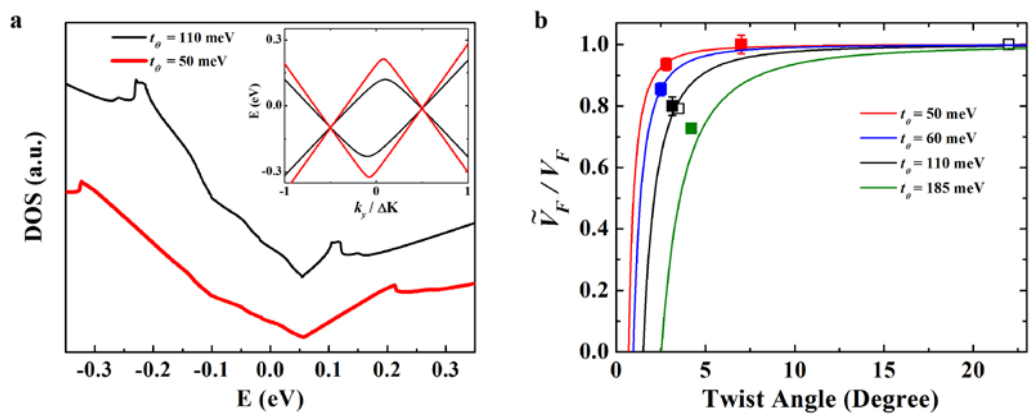

Figure 4

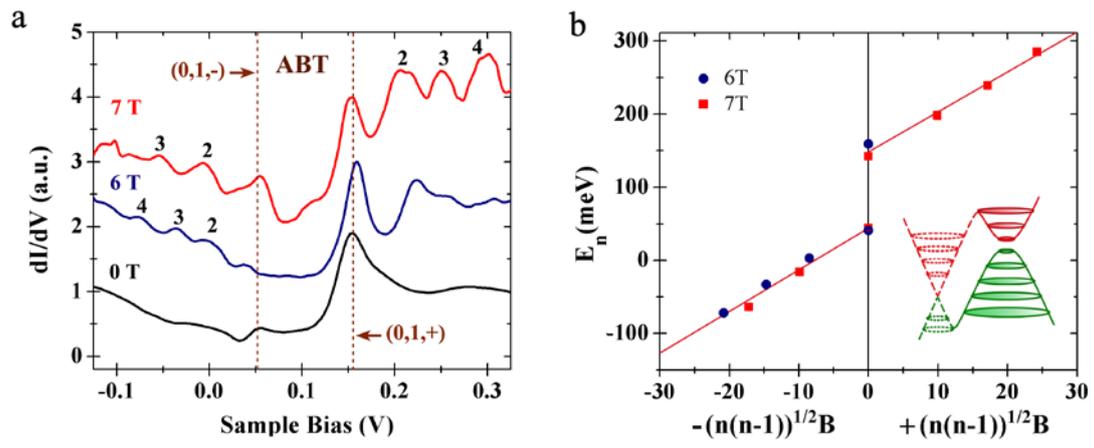

Figure 5